\documentstyle[12pt,a4]{article}

\def\title#1{\begin{center}{\Large{\bf #1}}
\end{center}}
\def\author#1{\begin{center} #1 \end{center}}
\def\address#1{\begin{center} {\it #1 } \end{center}}
\def\keywords#1{\begin{flushleft} {\large KEYWORDS:}
#1\end{flushleft}}

\def\preprint#1{
\hfill
\begin{flushright}
#1
\end{flushright}
}
\def\eline{\vfill
\begin{picture}(300,5)
\put(0,0){\line(1,0){280}}\end{picture}\\}

\def\eqn#1{\begin{eqnarray}#1\end{eqnarray}}
\def\non{\nonumber}

\def\ga{\gamma}		
\def\be{\beta}
\def\al{\alpha}

\def\de{\delta}		\def\De{\Delta}

\def\descriptionlabel#1{\bf #1\hfil}
\def\description{\list{}{%
\labelwidth=\leftmargin
\advance \labelwidth by -\labelsep
\let \makelabel=\descriptionlabel}}

\begin{document}

\preprint{OUCMT-94-10}
\title{Coupled-Map Modeling of One-Dimensional \\Traffic Flow${}^{\dagger}$}
\author{Satoshi Y{\footnotesize UKAWA}${}^{\ddag}$  and Macoto
K{\footnotesize IKUCHI}${}^{\ddag \ddag}$
}
\address{Department of Physics, Osaka University,
Toyonaka 560}

\begin{abstract}

We propose a new model of one-dimensional traffic flow using
a coupled map lattice.
In the model, each vehicle is assigned a map and changes its
velocity according
to it. A single map is designed so as to represent the motion of a vehicle
properly, and the maps are coupled to each other through
the headway distance.
By simulating the model, we obtain a plot of the flow against
the concentration similar to the
observed data in real traffic flows. Realistic traffic jam regions are
observed in space-time trajectories.

\end{abstract}

\eline
$\dagger$ to appear in J. Phys. Soc. Jpn.{\bf 64}(1995) January.\\
\ddag {\small e-mail: yuk@phys.sci.osaka-u.ac.jp}\\
\ddag \ddag {\small e-mail: kikuchi@phys.sci.osaka-u.ac.jp}\\

\newpage

Theoretical study of traffic flow has been made from two points of view,
macroscopic and microscopic. In the former, traffic flow is
regarded as a compressible fluid and fluid-dynamical ideas are
used.${}^{1)}$
In the latter, on the other hand, vehicles are treated individually.${}^{2-4)}$
Car-following models${}^{2,4)}$ are typical microscopic models which
are described by simultaneous differential equations.
These models have been used chiefly for rather high vehicle concentration.
For low concentration, they have some difficulties in realizing real
traffic flow;
since the velocity of vehicles is determined only by the relative
velocity and the mutual headway distance, the models exhibit some
unphysical behaviors, such as negative velocity and
limitless acceleration. Moreover, collisions of two vehicles
occur easily. They fail to describe the difference between free flow
and congested flow.
Recently a new car-following-type model has been proposed by Bando et
al.,${}^{5)}$
which seems to overcome some of these difficulties.
While these models are constructed in continuous space-time,
cellular automata (CA),${}^{6)}$ which
are discrete both in space and time, recently came into use for
modeling traffic flow.${}^{7-13)}$
Although CA can reproduce some properties of traffic flows,
they are as yet too simple as models of real traffic.

In this letter, we propose a new microscopic model of one-dimensional
traffic flow using the coupled map lattice (CML) idea.${}^{14)}$
The present model is constructed in continuous space, but in discrete time.
To make the model concrete, we start by considering the motion of a
single vehicle.
In real traffic flow, each vehicle has its own preferred velocity $v^F$
and adjusts its velocity so as to fit the preferred one.
Once the velocity reaches $v^F$, it fluctuates around $ v^F$.
For simplicity, when the current velocity is far from $v^F$, we assume
constant acceleration or constant deceleration.
The adjustment
mechanism and the fluctuations of velocity are
realized by the following map assigned to the vehicle:
\eqn{
v^{t+1} = F(v^t) \equiv  \ga v^t + \beta \tanh \left(
\frac{v^F-v^t}{\delta} \right) + \epsilon,
\label{eq:chaoticmap}
}
where $v^{t}$ and $v^F$ are the velocity of the vehicle at
time $t$ and the preferred velocity of the vehicle,
respectively, and $ \beta, \gamma, \delta,$ and $ \epsilon $ are the
parameters. In this model, the velocity is defined as the distance
traveled in a unit time step.
When $\ga = 1$ and $\be \ne 0 $, this map
expresses constant acceleration and deceleration for velocity far
from $v^F$.
We, however, take $\ga$ slightly greater or less than 1, since
the map then becomes chaotic, and acceleration and deceleration are
still approximately constant.
As a result, fluctuations in velocity are naturally introduced
to the model through deterministic chaos. It is worth noting that the
map $F(v)$ is closely related to that used in chaotic
neural network models.${}^{15)}$
The meaning of each parameter is as follows: $\delta $ determines the
fluctuation of
velocity around $v^F$; $\be$
represents the magnitude of acceleration and deceleration, and
difference in their magnitude is given by $\epsilon $.
Throughout this paper, we use the values $\be = 0.6, \ga=1.001,
\de = 0.1,$ and $ \epsilon = 0.1$.
The motion of one vehicle driven by $F(v)$ is given as a space-time
trajectory in Fig.~1. Weak fluctuations caused by chaos
are clearly seen.

Next, we consider the case where the number of vehicles is more than one.
In this situation, another deceleration mechanism is necessary to
avoid collision. In fact, we assign a deceleration map
to each vehicle.
In real traffic flows, a vehicle changes its velocity according to,
for example, the current velocity, the headway distance, and the
relative velocity with another vehicle, but here we assume that
the deceleration is
dominated by the headway distance from the nearest vehicle ahead.
Therefore the deceleration map determines the subsequent
velocity from the headway distance.
We introduce two deceleration processes, i.e., a sudden braking process and a
slowing-down process.
We call the model which includes the sudden braking process model A,
and the one which includes the slowing-down process model B.
The sudden braking process is very simple. Suppose a vehicle has the
velocity $v_0$
and another vehicle exists at the headway distance
$ \Delta x -l $, where $ \De x $ is the head-to-head distance and $ l
$ is the vehicle length;
if $\Delta x - l < v_0 $,
the vehicle changes its velocity to $ \Delta x -l $.
Consequently, a collision of vehicles is avoided.
In model B, the deceleration process is more complex; it
is described by a map from the headway distance to the
velocity defined as follows:
\eqn{
v_{i}^{t+1} = G(\De x_i^t,v_i^t)  & \equiv  &
\frac{F(v_i^{t})-v_{i}^t}{(\alpha-1) v_{i}^t}
(\De x_i^t-l-v_i^t) + v_i^t, \non \\
 & & v_i^t \le \De x_i^t - l \le \al v_i^t,
\label{eq:decmap}
}
where $\De x_i^t =x_{i+1}^{t}-x_{i}^{t}$ with positions of the {\it
i}th vehicle and the ({\it i}+1)th vehicle at time $t$ denoted by
$x_{i+1}^t$ and $x_i^t$, respectively.
The parameter $\al $ represents the range within which the vehicle uses the
deceleration map $G( \De x, v^t )$. If the headway distance is
less than $ \al v^t $, the deceleration map $G( \De x, v^t )$ is used
instead of $F(v)$.
The map gives $\De x_i^t -l$
for $\De x_i^t -l=v_i^t$ and $F(v_i^t)$ for $\De
x_i^t-l= \al v_i^t$.
Therefore, it connects the free-motion map $F(v)$
with the sudden braking process.
The velocity given by this map will be either greater or
less than the current velocity, but it is always lower than the value
given by the free-motion map
$F(v_i^t)$.

We simulate the two models on a circular road.
The positions and the velocity of vehicles are updated parallelly by the
following process. First, the headway distance is measured for all vehicles.
Then the vehicles move simultaneously according to the headway
distance. If the headway
distance is larger than the current velocity, the value of velocity
is added to the current position. If the headway distance is less than the
current velocity, on the other hand, the headway distance is added
to the position.
Next we determine the subsequent velocities for all the
vehicles according to the maps.
These processes constitute one time step.
We define some quantities observed in the
simulation: the average velocity $ \langle v \rangle $ is defined
as the distance traveled per time step
per vehicle, and the traffic flow as $ \rho \times  \langle v \rangle
$ using the vehicle concentration $\rho$. In the actual simulation,
the distance is measured in the units of car length.
The initial velocity and the
perferred velocity are distributed uniformly in [2.0,4.0].
Initial positions are randomly chosen.
In the simulation of model B, we adopt $\al = 4.0$.

In Fig.~2, we show the plot of the flow against the concentration
for both models A and B.
This diagram shows similar properties to those observed in real
traffic:${}^{16)}$
at low concentration, the flow increases almost linearly; after a
clear peak, it decreases slowly for higher concentration.
In Figs.~3(a)-3(c), typical space-time trajectories of model B for
several concentrations are shown.
Figure~3(a) corresponds to the concentration $\rho = 0.10 $, which is
lower than the peak concentration in Fig.~2.
In this figure, clustering of the vehicles is seen.
The cluster is led by the vehicle whose perferred velocity is
lowest. This vehicle moves according to the free-motion map $F(v)$.
It is seen that the fluctuation in its trajectory propagates to the
following vehicles.
As a result, weak shock waves form from time to time.
Figure~3(b) corresponds to the concentration $\rho = 0.30$
, which is higher than the the peak concentration.
A jam region within which vehicles stop is seen. We call this type
of jam a hard jam. The hard jam region moves backward. In real traffic
flow, the hard jam is frequently observed.
Figure~3(c) corresponds to $ \rho = 0.20 $, which is close to the
peak concentration.
A jam region where vehicles move
more slowly than in other regions is clearly seen.
We call this type of jam a soft jam. Its front moves slowly with positive
velocity.
A soft jam with negative velocity and hard jam are also found at
the same concentration depending on the initial configuration.
The soft jam with negative velocity is
also observed in real traffic flows.
The behavior of model A is similar to that of model B.

To summarize, we proposed a new model of one-dimensional
traffic flow using the CML idea.
The present model has several advantages over the traditional
car-following
model: the fluctuations of a single vehicle are taken into account;
there is no collision of vehicles because of the discretization of time.
In the space-time trajectory, the motion of vehicles escaping from
the jam region seems very similar to that in real traffic.
The motion of vehicles catching up to the jam
region, on the other hand, is less satisfactory.
We are currently trying to improve the modeling of the deceleration process.
Detailed study on the model including the effect of chaos is now in
progress.

We are grateful to Y. Akutsu, K. Tokita, M. Yamashita, S. Tadaki,
M. Bando, K. Hasebe, and M. Kawahara for valuable
discussions. The work was supported in part by a research grant from
The Japan Science Society.

\newpage
{\large {\bf References}}
\begin{description}

\item[1)] M. J. Lighthill and G. B. Whitham: Proc. Roy. Soc. {\bf
A 229} (1955) 317.

\item[2)] R. Herman, E. W. Montroll, R. B. Potts and
R. W. Rothery: Oper. Res. {\bf 7} (1959) 86.

\item[3)] I. Prigogine and F. C. Andrews: Oper. Res. {\bf
8} (1960) 789.

\item[4)] D. C. Gazis, R. Herman and R. W. Rothery:
Oper. Res. {\bf 9} (1961) 545.

\item[5)] M. Bando, K. Hasebe, A. Nakayama, A. Shibata and
Y. Sugiyama: Japan J. Indust. Appl. Math. {\bf 11} (1994) 203.

\item[6)] S. Wolfram: Rev. Mod. Phys. {\bf 55} (1983) 601.

\item[7)] O. Biham, A. A. Middleton and D. Levine:
Phys. Rev. {\bf A46} (1992) 6124.

\item[8)] J. A. Cuesta, F. C. Mart\' \i nez, J. M. Molera and
A. S\'anchez: Phys. Rev. {\bf E48} (1993) R4175.

\item[9)] T. Nagatani: Phys. Rev. {\bf E48} (1993) 3290;
J. Phys. Soc. Jpn. {\bf 63} (1994) 52; J. Phys. Soc. Jpn. {\bf 63}
(1994) 1228.

\item[10)] S. Tadaki and M. Kikuchi: to appear in Phys. Rev. E.

\item[11)] S. Yukawa, M. Kikuchi and S. Tadaki:
J. Phys. Soc. Jpn. {\bf 63} (1994) 3609.

\item[12)] K. Nagel and M. Schreckenberg: J. Phys. I France {\bf
2} (1992) 2221.

\item[13)] M. Takayasu and H. Takayasu: Fractals {\bf 1} (1993) 860.

\item[14)] K. Kaneko ed.: {\it Theory and Application of Coupled Map
Lattices} (Wiley, New York, 1993) and references therein.

\item[15)] H. Nozawa: CHAOS {\bf 2} (1992) 377.

\item[16)] W. Leutzbach: {\it Introduction to the Theory of
Traffic Flow }(Springer-Verlag, Berlin, 1988).

\end{description}

\newpage
{\large Figure Captions}
\begin{description}

\item[Fig.~1.] The motion of a vehicle moving freely.
The initial velocity and the preferred velocity are taken to be
3.0. The system length is 500.
We show the trajectory of 20 time steps after discarding 500 steps.

\item[Fig.~2.] The concentration
vs flow of models A and B.
The initial velocity and the perferred
velocity of
the vehicles are uniformly distributed in the range $[2.0,4.0]$.
The system length is 500.
We show the average over 100 time steps after discarding 500 steps.

\item[Fig.~3~(a).] The space-time trajectory of model B.
The initial velocity and the favorite velocity of the vehicles are
uniformly distributed in the range $[2.0,4.0]$. The system length is
100. One hundred time steps are shown after discarding 500 time steps.
The number of vehicles is 10.

\item[~~~~~~~(b).] The space-time trajectory of model B.
The number of the vehicles is 30. The other
parameters are the same as for~(a).

\item[~~~~~~~(c).] The space-time trajectory of model B.
The number of vehicles is 20.
The other parameters are the same as for~(a).

\end{description}

\end{document}